\documentclass[a4paper,twocolumn,aps,showpacs,showkeys,preprintnumbers,amsmath,amssymb]{revtex4}
\usepackage{graphics}

\usepackage{graphicx}
\usepackage{dcolumn}
\usepackage{bm}
\usepackage{amssymb}

\newcommand{\be}{\begin{equation}}
\newcommand{\ee}{\end{equation}}
\newcommand{\bea}{\begin{eqnarray}}
\newcommand{\eea}{\end{eqnarray}}
\newcommand{\bean}{\begin{eqnarray*}}
\newcommand{\eean}{\end{eqnarray*}}
\newcommand{\bi}{\begin{itemize}}
\newcommand{\ei}{\end{itemize}}
\newcommand{\bdm}{\begin{displaymath}}
\newcommand{\edm}{\end{displaymath}}
\newcommand{\etal}{{\it et al. }}
\newcommand{\ie}{{\it i.e. }}
\newcommand{\eg}{{\it e.g. }}
\newcommand{\cf}{{\it c.f. }}

\begin{document}
\preprint{
\hfill
\begin{minipage}[t]{3in}
\begin{flushright}
\vspace{0.0in}
OUTP-0605P
\end{flushright}
\end{minipage}
}

\title{Constraints on UED KK-neutrino dark matter from magnetic dipole moments}

\author{Thomas Flacke}
\email{t.flacke1@physics.ox.ac.uk}
\affiliation{Rudolf Peierls Centre for Theoretical Physics, University of Oxford,
1 Keble Road, Oxford OX1 3NP, United Kingdom}

\author{David W.~Maybury}
\email{d.maybury1@physics.ox.ac.uk}
\affiliation{Rudolf Peierls Centre for Theoretical Physics, University of Oxford,
1 Keble Road, Oxford OX1 3NP, United Kingdom}

\date{January 24, 2006}
\begin{abstract}
Generically, universal extra dimension (UED) extensions of the standard model predict the stability of the lightest Kaluza-Klein (KK) particle and hence provide a dark matter candidate. For UED scenarios with one extra dimension, we model-independently determine the size of the induced dimension-five magnetic dipole moment of the KK-neutrino, $\nu^{(1)}$. We show that current observational bounds on the interactions of dipole dark matter place constraints on UED models with KK-neutrino dark matter.
\end{abstract}


\maketitle


\section{Introduction}

While the standard model proves an excellent framework for fundamental interactions at energy scales up to at least the sub TeV range, it nevertheless leaves a number of fundamental problems. Theoretically, one of the most outstanding puzzles centers on the origin and mechanism of electroweak symmetry breaking, and the quantum mechanical stability of the hierarchy generated between the electroweak scale and the Planck scale. In addition, recent astrophysical observations \cite{DarkMatter} concord with $0.094 < \Omega_{CDM}h^2 < 0.129$, indicating the presence of cold non-baryonic dark matter as the principle form of matter in the Universe, of which the standard model provides no explanation. The most popular candidate for dark matter assumes a non-standard model, stable, electrically neutral, and weakly interacting particle -- the WIMP hypothesis. Clearly, from both a theoretical and phenomenological perspective, the standard model requires extension. In this letter we wish to explore the consequences of the universal extra dimension (UED) \cite{Appelquist:2000nn} extension of the standard model.

In the UED scenario, all standard model particles can freely propagate in the bulk of one or more extra dimensions and thus each standard model particle is associated with a Kaluza-Klein (KK) tower of states. Each state in the KK tower has the same spin as its standard model counterpart. An important consequence of UED models concerns the existence of a conserved discrete symmetry, KK-parity, which guarantees the stability of the lightest KK particle (LKP) and thus provides a dark matter candidate. Suitable thermal relic dark matter candidates that have been studied extensively \cite{Servant:2002aq} include the first KK-excitations of the hypercharge boson, the photon, and the neutrino \ie $B^{(1)}$, $\gamma^{(1)}$, or $\nu^{(1)}$.

The tree-level mass spectrum of the KK-excitations of UED models reveals a nearly degenerate spectrum. As an example, a UED model with one extra-dimension compactified on an $S_1/Z_2$ orbifold of radius $R$, leads to the tree level mass relation 
\be\label{mass} m^{(n)} = \sqrt{(n/R)^2 + (m^{(0)})^2} 
\ee 
for the n-th KK mode, where $m^{(0)}$ constitutes the zero-mode mass (\ie the standard model particle value). Quantum corrections typically dominate over zero mode level contributions and therefore the resulting mass spectrum depends crucially on radiative effects. In general, a moderately split UED mass spectrum \cite{Cheng:2002ej} develops. In the minimal UED model (MUED) \cite{Cheng:2002ej}, one-loop calculations suggest that the LKP is well approximated by the KK hypercharge boson, $B^{(1)}$, and numerous studies have examined the thermal production and prospects of direct and indirect detection of $B^{(1)}$ and $\gamma^{(1)}$ LKP dark matter \cite{Servant:2002aq,KKDMcollection}. 

However, the non-renormalizability of UED models imply the existence of an ultraviolet cut-off, typically of the order of a few tens of TeV, at which point the model requires UV completion. As such, UED models must be regarded as an effective theory. The presence of incalculable boundary terms arising from the UV complete theory can potentially change the mass spectrum, resulting in different LKP candidates. Recent studies of the relic density of $B^{(1)}$ LKP dark matter with the full MUED spectrum \cite{coannmat,coannkribs} reveal substantial observational tension with constraints from electroweak precision data \cite{Flacke:2005hb}. Thus, non-minimal models with brane-localized terms appear as a likely alternative if UED models are to provide a successful phenomenology. Furthermore, model independent studies \cite{Servant:2002aq} show that $B^{(1)}$, $\gamma^{(1)}$, and $\nu^{(1)}$ can all be thermally produced with abundances sufficient to provide the dark matter. Constraints on minimal UED models from limits on weak neutral current nucleon-$\nu^{(1)}$ elastic scattering in direct searches, together with thermal dark matter production mechanisms, disfavour $\nu^{(1)}$ dark matter \cite{Servant:2002hb}. However, given the need for possible new non-minimal interactions, we consider further consequences of KK-neutrino dark matter model-independently.

While there exists compelling evidence for dark matter in the form of WIMPS, there also exists strong constraints on possible electromagnetic interactions of dark matter, even in the limit of complete charge neutrality. A neutral Dirac fermion can posses both a permanent magnetic dipole moment, $\mu$, and a permanent electric dipole moment, $d$, arising from the dimension-five operator, 
\be 
\mathcal{L}_{D} = \frac{i}{2} \bar f
\sigma_{\mu\nu}\left(\mu +\gamma_5 d \right)f F^{\mu\nu}. 
\ee
While the presence of the magnetic dipole moment does not violate any discrete symmetries, the electric dipole moment requires the violation of parity and CP. Severe constraints exist on the dipole moments of $\sim 1$ TeV dark matter WIMPS \cite{Sigurdson:2004zp}. (We are aware that the authors of \cite{Sigurdson:2004zp} are currently revising their estimates and, as a result, the strength of the constraints in \cite{Sigurdson:2004zp} may change substantially \cite{private}.) Thus, if a model predicts Dirac fermionic dark matter, it is important to determine the strength of the induced dipole moment.

Since the KK-neutrino of UED models is a Dirac fermion from the 4-dimensional perspective, UED models that assume KK-neutrino dark matter are constrained by the strength of the induced dipole operator.

In this letter we derive model independent bounds on KK-neutrino dark matter by examining the induced dipole moment and comparing the predictions with the current observational bound. Section II provides a brief review on the properties of KK-fermions in UED models along with a discussion on dipole moments relevant to the calculation presented in section III. Finally, in section IV we present our conclusions.

\section{KK-neutrino LKP and induced dipole moments}

In UED models, standard model fields become identified with the zero modes of KK towers of states once the extra dimensions are integrated out of the theory. For concreteness, we will restrict our discussion to five dimensional UED models compactified on an $S_1/Z_2$ orbifold. Five dimensional theories do not posses a chirality condition since $\gamma_5$ becomes part of the five dimensional Clifford Algebra. Thus, in order to arrive at a chiral theory at the zero mode level, we require one five dimensional \emph{Dirac} spinor for every \emph{Weyl} spinor of the standard model. By use of the orbifold boundary conditions half the number of states project out of the spectrum leaving a zero mode level chiral theory. In the case of the lepton doublet, the decomposition of the corresponding five dimensional Dirac spinor reads (\cf \eg
\cite{Appelquist:2000nn}),
\be \label{fermiondecomp}
\begin{split}
\hat{\mathcal{L}}(x^{\mu},y) = \frac{1}{\sqrt{\pi R}}
P_L\mathcal{L}(x^\mu) +  \sqrt{\frac{2}{\pi
R}}\sum_{n=1}\left[P_L\mathcal{L}(x^\mu)\cos(\frac{ny}{R})\right.&\\
  \left.+P_R\mathcal{L}(x^\mu)\sin(\frac{ny}{R})\right]&\\
\end{split}
\ee
where $\hat{\mathcal{L}}$ denotes the five dimensional Dirac spinor, $\mathcal{L}$ denotes a four dimensional \emph{4
component} spinor with $P_{L,R} = (1\pm\gamma_5)/2$, and $\hat{\mathcal{L}}$ is associated with the (4D left-handed) lepton doublet via the identification $P_L\mathcal{L}(x^\mu)\leftrightarrow (\nu_L,e_L)$. We see from eq(\ref{fermiondecomp}) that $\hat{\mathcal{L}}$ contains a purely left-handed zero mode while all higher KK modes contain both chiralities. Therefore, the non-zero mode KK-level fermions appear as Dirac spinors from the 4-dimensional perspective. Specifically, the KK-neutrino charged under the weak $SU(2)$ appears with both chiralities.

Translational invariance in the fifth direction implies (once the extra dimension becomes integrated out) that a UED model compactified solely on $S_1$ conserves KK-excitation-number in every vertex. The orbifold reduces the conserved KK-number to a discrete $Z_2$ symmetry, called KK-parity. The conservation of KK-parity in every vertex implies the stability of the lightest KK-particle.

Since the fermions receive mass through Yukawa couplings after electroweak symmetry breaking and through the KK-level expansion itself, the theory, in general, requires a unitary transformation to connect mass and gauge eigenstates. For example, in the lepton sector at the $j$th KK-level we have,
\be 
\left(\begin{array}{c} \mathcal{E}^j \\ \mathcal{L}^j \end{array} \right) =
\left(\begin{array}{cc}-\gamma_5\cos\alpha_j & \sin\alpha_j\\
\gamma_5\sin\alpha_j & \cos\alpha_j
\end{array}\right)\left(\begin{array}{c} \mathcal{E}^{\prime j} \\
\mathcal{L}^{\prime j} \end{array} \right) 
\ee 
where $\mathcal{E}^j$ and $\mathcal{L}^j$ denote the $j^{\mbox{th}}$ KK mode of the lepton $SU(2)$ singlet and doublet respectively, and where \be \tan (2\alpha_j) = \frac{m_l^{(0)}}{j/R}. \ee As the lepton masses are small compared to $j/R$, we ignore the effects of the mixing matrix for the remainder of this letter. Furthermore, we ignore the effects of lepton flavour violation and neutrino mixing.

A Dirac fermion can posses a dipole moment, derived from the transition amplitude (\cf \eg \cite{neutrinobook}), 
\be 
T =
-i\epsilon^\mu q^\nu \bar f (p^\prime) \sigma_{\mu\nu} \left(F_2 + G_2
\gamma_5\right) f(p) 
\ee 
where $q=p^\prime -p$. The magnetic moment is defined by $\mu = F_2(0)$ while the electric dipole is defined by $ d = G_2(0)$. At low energies compared to the mass of the particle, the photon does not distinguish between $\mu$ or $d$ provided that one ignores other time-reversal violating observables. We will focus on the limits established on $\mu$ throughout. On dimensional grounds, we naively expect the induced magnetic dipole moment to scale as, 
\be 
\mu \lesssim e \frac{M_{\nu^{(1)}}}{R^{-2}} \simeq \frac{e}{M_{\nu^{(1)}}} = 1.022\times
10^{-6}\mu_B\left(\frac{\mathrm{TeV}}{M_{\nu^{(1)}}}\right) 
\ee
where $M_{\nu^{(1)}}$ denotes the mass of KK-neutrino, $\nu^{(1)}$, and $R$ indicates the radius of compactification.

\section{Calculation}

\begin{figure}[h]
     \begin{center}
     	\vspace{5pt}
 	\hspace{-100pt}{\includegraphics[height=120pt,width=120pt]{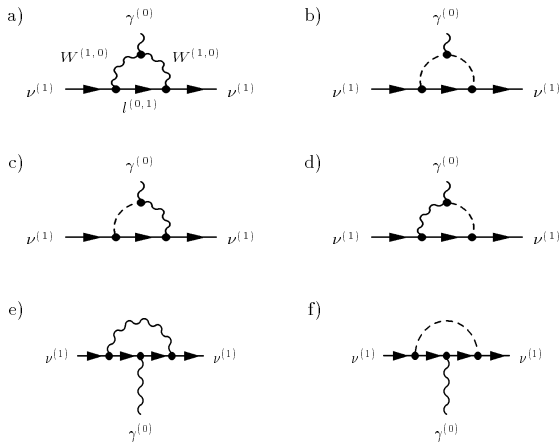}}
     	\vspace{45pt}
    	\caption{Magnetic dipole moment inducing one loop Feynman graphs for
             the first KK-excitation of the neutrino contained in the $\mathrm{SU(2)}_L$ doublet in UED models.
             KK-number assignments are denoted in graph a), to be understood analogously in graphs b) - f).
             Graphs with dashed lines denote the Goldstone modes along their
             KK-excitations, and the KK-excitation of the Higgs scalar. The Feynman rules are taken from \cite{Buras:2002ej}.}
	\label{Figgraphs}     
     \end{center}
\end{figure}

The KK-neutrino develops a magnetic dipole moment through the diagrams tabulated in figure \ref{Figgraphs}.
In general, the entire Kaluza-Klein tower of states participate, however we estimate the leading order effect by considering only the first level KK excitations. Furthermore, we restrict our calculation to KK-number conserving graphs since KK-number violation proceeds with volume suppression. For simplicity, we ignore flavour violation in the lepton sector and we assume that the lightest KK-neutrino is $\nu_e^{(1)}$. Relaxing these assumptions will not significantly alter our conclusions.

As we make no assumptions on the exact UED spectrum, we consider the mass of the KK-$W$ ($W^{(1)}$) and the KK-electron ($e^{(1)}$) as free parameters. In our numerical calculations we do not consider a KK-electron/KK-neutrino mass difference in excess of 5\% as any substantial splitting will lead to unacceptably large contributions to the $T$ parameter.

\begin{figure}[ht!]
   \newlength{\picwidtha}
   \setlength{\picwidtha}{3.5in}
   \begin{center}
       \resizebox{\picwidtha}{!}{\includegraphics{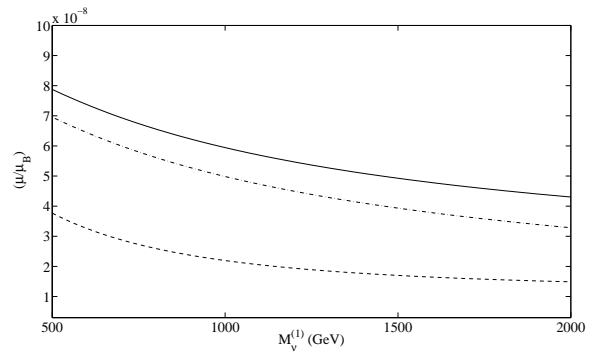}}
   \end{center}
   \caption{Magnetic moment of the KK-neutrino $\nu^{(1)}$ vs.
            KK dark matter mass $M_{\nu^{(1)}}$ for varying $M_{e^{(1)}}$, and $M_{W^{(1)}}$.
            Upper line (solid): $M_{e^{(1)}}$-$M_{\nu^{(1)}}$ degeneracy with tree-level $M_{W^{(1)}}$  (\cf analytical result eq(\ref{dipole_result})).
            Middle line (dashed-dotted): 5\% splitting of $M_{e^{(1)}} / M_{\nu^{(1)}}$ with with tree-level $M_{W^{(1)}}$.
            Lower line (dashed): $M_{e^{(1)}}$-$M_{\nu^{(1)}}$ degeneracy with 5\% splitting of $M_{W^{(1)}} / M_{\nu^{(1)}}$}
	    \label{plot_1}
\end{figure}

The relevant UED Feynman rules are listed in \cite{Buras:2002ej} and we calculate in the Feynman-'t Hooft gauge. In the limit of exact $M_{\nu^{(1)}}$-$M_{e^{(1)}}$ degeneracy and where the effects of Yukawa couplings are ignored, we arrive at the semi-analytic result,
\be 
\label{dipole_result} 
\begin{split}
\mu =   & \frac{eg^2}{(4\pi)^2}\frac{1}{M_{\nu^{(1)}}} \times\\
	&\left\{\frac{3}{2}\ln (\epsilon)+r+\frac{7}{2}+\frac{1}{2r}-\frac{5}{2}(r-1)\ln \left(\frac{r}{r-1}\right)\right.\\
	&\left. -(r-1)^2\ln \left(\frac{r}{r-1}\right) + \mathcal{O}(\sqrt{\epsilon})\right\}	
\end{split}
\ee 
with the approximation $\epsilon\equiv M_{W^{(0)}}^2/M_{\nu^{(1)}}^2\ll 1$ and $r\equiv M_{W^{(1)}}^2/M_{\nu^{(1)}}^2\simeq 1$.
Numerically, we find agreement with our semi-analytical result as seen in figure \ref{plot_1}.

In figure \ref{plot_1} we display the result of the dipole moment as a function of the KK-neutrino mass. The upper curve illustrates the magnetic dipole moment with exact degeneracy $M_{e^{(1)}}$-$M_{\nu^{(1)}}$ while holding the KK-$W$ mass at its tree level value. The middle curve plots the effect of a KK-electron 5\% heavier than the KK-neutrino. This has an $\mathcal{O}(1)$ effect on the dipole moment. The predicted value of the dipole moment exceeds the current upper bound for a KK-neutrino mass $M_{\nu^{(1)}}\sim 1 \hspace{1mm}\mathrm{TeV}$ by more than five orders of magnitude \cite{Sigurdson:2004zp}. The lower curve displays the effect of maintaining $M_{\nu^{(1)}}$-$M_{e^{(1)}}$ degeneracy while varying the KK-$W$ mass. Allowing the mass difference, $M_{W^{(1)}}$-$M_{\nu^{(1)}}$, to vary by up to 5\% has at most an $\mathcal{O}(10)$ effect.

We should note that the calculation presented above determines only the radiatively induced part of the KK-neutrino magnetic dipole moment. The presence of boundary terms or effects arising from the UV complete theory may also contribute a non-renormalizable dimension-five dipole operator which, a priori, may be of the same order as the radiative part itself.

\section{Conclusion}

UED models have attracted attention as a possible extension to the standard model. A particular appealing feature of the model class centers on the existence of plausible dark matter candidates as the result of KK-parity conservation. While the minimal UED model suggests $B^{(1)}$ dark matter \cite{Cheng:2002ej}, detailed studies of the relic abundance in the minimal UED model \cite{coannmat,coannkribs, Matsumoto:2005uh} in combination with electroweak precision constraints \cite{Flacke:2005hb} show strong observational tension, and thus provide motivation for new possible UED model building avenues. The need for non-minimality has been reported \cite{Hewett:2004py}. Extensions of the MUED scenario by incalculable boundary terms arising from the UV completion of the model can lead to a different LKP and therefore different possible dark matter candidates. We have taken a model independent approach, following \cite{Servant:2002aq}, and examined the consequences of UED KK-neutrino dark matter.

As the KK-neutrino is a Dirac fermion, UED models predict an induced KK-neutrino dipole moment. We find that the induced dipole moment, typically $\mu \lesssim 10^{-7} \mu_B$, strongly conflicts -- by over five orders of magnitude -- with the current observational bounds stated in \cite{Sigurdson:2004zp} for TeV scale dipole dark matter. We reiterate that the constraints provided by \cite{Sigurdson:2004zp} are currently under revision, and the strength of the stated bounds are expected to change \cite{private}. The constraints on magnetic dipole moments given in \cite{Sigurdson:2004zp} would provide the strongest limits on KK-neutrino dark matter. Even in the absence of the strong limits provided by \cite{Sigurdson:2004zp}, the bounds on dipole moments remain an important constraint on future model building. Not only will new, non-minimal models that predict KK-neutrino LKP need to circumvent the constaints provided by \cite{Servant:2002hb}, but will also have to evade the constraints provided by radiatively induced magnetic dipole moments, which are generically at least as large as the current experimental bounds. 

While we have restricted our discussion to five-dimensional models compactified on $S_1/Z_2$, we expect that the qualitative features carry over to UED models with multiple extra dimensions. We have also assumed the absence of any fine-tuning between the radiatively induced magnetic dipole moment and possible non-renormalizable dimension-five dipole operators arising from the UV complete theory or boundary terms.

Our results indicate that observational limits on dipole dark matter can place significant constraints on UED scenarios where the KK-neutrino is the LKP dark matter candidate.

\section{Acknowledgments} 

We would like to thank J.~March-Russell, G.~Starkman, B.~A.~Campbell, and K.~Sigurdson for useful discussions. DM wishes to acknowledge the support of the Natural Science and Engineering Research Council of Canada and the Canada-United Kingdom Millennium Research Fellowship. The work of TF is supported by ``Evangelisches Studienwerk Villigst e.V." and PPARC Grant No. PPA/S/S/2002/03540A. This work was also supported by the ``Quest for Unification" network, MRTN 2004-503369.

\end{document}